\documentstyle[a4]{article}
\def\abstract#1{\vskip 7mm 
        \begin{center}{\large Abstract}\par \smallskip
                \begin{minipage}[c]{12cm}
                        \small #1
                \end{minipage}
        \end{center}
}
\def\title#1{\begin{center}{\Large\bf #1}\end{center}}
\def\author#1{\vskip 5mm \begin{center}{#1}\end{center}}
\def\address#1{\begin{center}{\it #1}\end{center}}
\begin{document}

\title{ Perturbative analysis of non-singular cosmological model\footnote{
Talk given at YITP, Kyoto, 
in the VIIth Workshop on General Relativity and Gravitation (Oct.27-30, 1997).}}
\author{Shinsuke {\sc Kawai}\footnote{E-mail:kawai@phys.h.kyoto-u.ac.jp}}
\address{Graduate School of Human and Environmental Studies, Kyoto University,
Kyoto 606, Japan}
\author{Masa-aki {\sc Sakagami}\footnote{E-mail:sakagami@phys.h.kyoto-u.ac.jp}
 and Jiro {\sc Soda}\footnote{E-mail:jiro@phys.h.kyoto-u.ac.jp}}
\address{Department of Fundamental Sciences, FIHS, Kyoto University, Kyoto 606,
Japan}
\abstract{
A stability analysis is made for a non-singular pre-big-bang like 
cosmological model based on 1-loop corrected string effective action.
Its homogeneous and isotropic solution realizes non-singular
transition from de Sitter universe to Friedmann-like universe, via super
inflation phase. We are interested in whether the non-singular nature of the 
solution would be stable or not in more general inhomogeneous case. 
Perturbative analysis is made for scalar, vector, and tensor linear 
perturbations, and instability is found for tensor-type perturbation.
}

%%%%%%%%%%%%%%%%%%%%%%%%%%%%%%%%%%%%%%%
\section{Introduction}
%%%%%%%%%%%%%%%%%%%%%%%%%%%%%%%%%%%%%%%

One of the most remarkable predictions of General Relativity is the existence
of singularities. Particularly at the center of the black hole and in the very
early universe, one might interpret the singularities as a result of the 
breakdown of General Relativity. If so, the behavior of the system ``near''
the singularity should be explained by some more general theory, preferably
including GR as an effective theory in particular situation. In this sense, 
searches for singularity-free cosmological models have been made by many 
authors
\cite{nonsingular}. Recently non-singular cosmological models based on 
superstring theory\cite{sstr} were presented, and among these, we will here
investigate the model presented in \cite{rt94}, which is derived from the 
superstring effective action with  1-loop string correction \cite{art94}
\begin{equation}
{\cal S}=\int d^4x\sqrt{-g}\{\frac 12 R-\frac 14(D\Phi)^2-\frac 34 (D\sigma)^2
 +\frac 1{16} [\lambda_1 e^\Phi-\lambda_2\xi(\sigma)]R^2_{GB}\}
\label{eqn:ss1leAct}
\end{equation}
where $R$ ,$\Phi$ and $\sigma$ are the Ricci scalar curvature, the dilaton, and
the modulus field, respectively.
The Gauss-Bonnet curvature is 
\begin{equation}
R^2_{GB}=R^{\mu\nu\kappa\lambda}R_{\mu\nu\kappa\lambda}
 -4 R^{\mu\nu}R_{\mu\nu}+R^2,
\end{equation}
and $\xi(\sigma)$ is a function of modulus field expressed with Dedekind $\eta$
function:
\begin{equation}
\xi(\sigma)=-\ln[2e^\sigma\eta^4 (i e^\sigma)].
\label{eqn:ssxi}
\end{equation}
Since the non-singular nature essentially depends on the behavior of modulus
field $\sigma$, we will concentrate on the modulus field only and actually use 
the effective action appearing in \cite{rt94}:
\begin{equation}
{\cal S}=\int d^4x\sqrt{-g}\{\frac 12 R-\frac 12 (D\varphi)^2
 -\frac{\lambda}{16}\xi(\varphi) R_{GB}^2\}.
\label{eqn:rt94act}
\end{equation}
From this action we will derive the back ground equations of motion and see
the behavior of the solution in the next section. In the section 3, the
method of perturbation is explained and equations of motion for perturbative
variables are derived, and also the stability of this cosmological model is 
discussed. The results are summarized in the section 4.

%%%%%%%%%%%%%%%%%%%%%%%%%%%%%%%%%%%%%%%
\section{Model and the background solution}
%%%%%%%%%%%%%%%%%%%%%%%%%%%%%%%%%%%%%%%

We will assume the homogeneous and isotropic metric with conformal time $\eta$,
\begin{eqnarray}
ds^2&=&a(\eta)^2 (-d\eta^2+\gamma_{ij}dx^i dx^j),\\
\label{eqn:bgmetric}
\gamma_{ij}&=&\frac 1{1+\frac 14 {\cal K}(x^2+y^2+z^2)}\delta_{ij}
\end{eqnarray}
where ${\cal K}=0,+1,-1$ for flat, closed, open universe, respectively.
From the action (\ref{eqn:rt94act}) one can derive the equations of motion as
\begin{eqnarray}
&&\varphi'{}^2=6({\cal H}^2+{\cal K})(1-\frac \lambda{2a^2}{\cal H}\xi')
\label{eqn:bg1}\\
&&({\cal H}'+{\cal H}^2+{\cal K})(1-\frac \lambda{2a^2}{\cal H}\xi')
 +({\cal H}^2+{\cal K})(1-\frac \lambda{4a^2}\xi'')=0
\label{eqn:bg2}\\
&&\varphi''+2{\cal H}\varphi'+\frac {3\lambda}{2a^2}({\cal H}^2+{\cal K})
 {\cal H}'\xi_{,\varphi}=0
\label{eqn:bg3}
\end{eqnarray}
Here,the conformal Hubble parameter
\begin{equation}
{\cal H}:=\frac{a'}{a}
\label{eqn:cnfhubble}
\end{equation}
is used and prime ($'$) denotes differentiation with respect to conformal time
$\eta$. We denote the derivative with respect to the physical time $t$ as dot
($\dot{}$), and usual Hubble parameter as $H:=\frac{\dot a}{a}$.
Since the action (\ref{eqn:rt94act}) is invariant under the change of
$\varphi$'s sign, we can choose the initial condition of $\varphi$ so that
$\varphi$ increases in time (at least near the initial point). Using
\begin{equation}
\xi=\frac 12 \varphi^2
\label{eqn:2xi}
\end{equation}
instead of (\ref{eqn:ssxi}), which is a good approximation near $\varphi=0$, 
the solutions 
of the equations of motion which continue to Friedmann like phase ($H>0$,
$\dot H<0$) at $t=0$ are shown in the fig.1. In these solutions $\varphi$
increases monotonically, so larger $\varphi$ corresponds to later time. 
As can be seen from the figure, there are two classes of such solutions ---
singular and non-singular solutions. Solutions a and b in the fig.1 are 
singular at $\varphi=0$, whereas c, d, e and f continues beyond the initial
singularity. One may also notice that these non-singular solutions approaches
de Sitter like solution ($H\simeq const \neq 0$) as $\varphi \rightarrow
-\infty$, $t\rightarrow -\infty$. This is peculiar to the
form of the potential (\ref{eqn:2xi}). If we use (\ref{eqn:ssxi}), the behavior
of the solutions near $\varphi =0$ are quite similar to those shown in the 
fig.1, but $H$ finally approaches zero in the infinite past. Although the
asymptotic behavior is quite different, we will use (\ref{eqn:2xi}) as $\xi$
in the actual numerical calculation since it is simple, and we are only
interested in the behavior of the solution near the singularity.
\input epsf.sty
\begin{figure}
% \vspace*{0cm}
% \hspace*{0cm}
\epsfxsize=12cm
\epsfysize=9cm
\epsffile{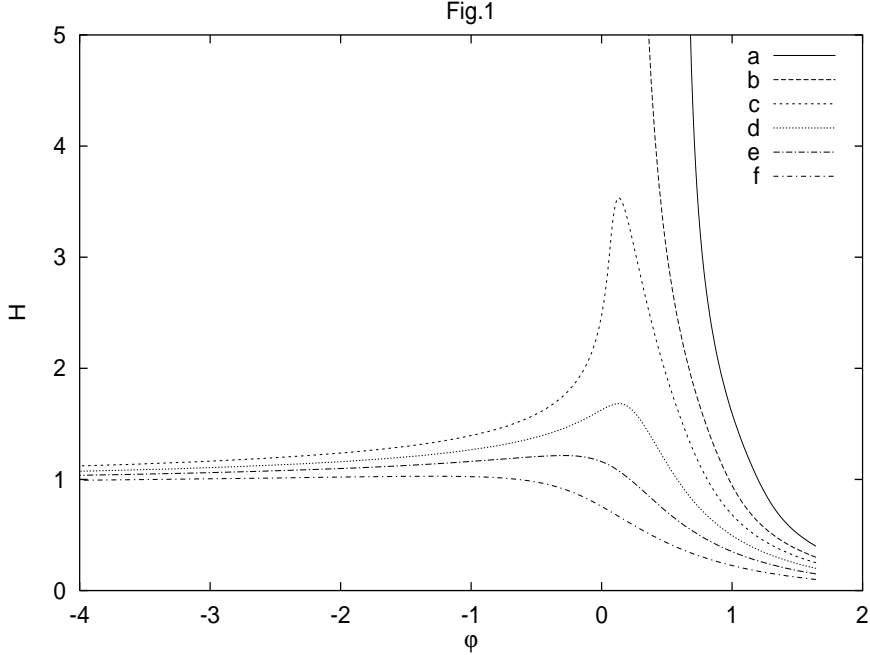}
\caption{Behavior of Hubble parameter versus scalar field $\varphi$. 
Time flows from left to right since in these solutions $\dot\varphi$ is always
positive. There are two classes of solutions: singular solutions (a and b), 
and non-singular solutions (c,d,e,f). We
can see that the initial singularity is avoided when $\varphi$ crosses zero.
For detailed discussions see \cite{rt94}.
}
\end{figure}

%%%%%%%%%%%%%%%%%%%%%%%%%%%%%%%%%%%%%%%
\section{Perturbative analysis}
%%%%%%%%%%%%%%%%%%%%%%%%%%%%%%%%%%%%%%%
\subsection{Method of perturbation}

 We will consider perturbation of our model(\ref{eqn:rt94act}) to analyze the 
stability of the non-singular nature. To avoid the gauge ambiguity, we here 
use the gauge-invariant perturbation method \cite{gip}. The metric is 
decomposed into the back ground part and the perturbed part, and the 
perturbed part into scalar-, vector-, and tensor-perturbation part:
\begin{equation}
g_{\mu\nu}={}^{(0)}g_{\mu\nu}+\delta^S g_{\mu\nu}
 +\delta^V g_{\mu\nu}+\delta^T g_{\mu\nu}.
\end{equation}
Also, the scalar field $\varphi$ (or the potential $\xi$) is decomposed into
its background and perturbed part:
\begin{eqnarray}
\varphi &=& {}^{(0)}\varphi+\delta\varphi,\nonumber\\
\xi &=& {}^{(0)}\xi+\delta\xi.
\end{eqnarray}
Change of perturbed values under gauge transformations is given by Lie 
derivatives as:
\begin{equation}
\delta Q \rightarrow \delta Q + \pounds_X {}^{(0)}Q.
\end{equation}
We will introduce such combinations of perturbed variables that the effects of 
gauge transformation cancel. These gauge invariant variables are used to 
describe the dynamics of the perturbation. Equations of motion for perturbed 
variables are obtained as perturbed Einstein-like equation:
\begin{equation}
\delta G^\mu{}_\nu=\delta T^\mu{}_\nu+\frac \lambda 2 \delta K^\mu{}_\nu.
\label{eqn:pertEin}
\end{equation}
Each term is also decomposed into scalar, vector, and tensor part.
Perturbed Klein-Gordon equation is not necessary since it is derived from
(\ref{eqn:pertEin}).
\subsection{Scalar part}
Scalar part of the perturbed metric is written in terms of 4 scalar functions
$\phi$, $\psi$, $B$, and $E$,
\begin{equation}
 \delta^S g_{\mu\nu}=a^2(\eta)\left(
   \begin{array}{cc}
     -2\phi & B_{|i}\\
     B_{|j} & -2(\psi\gamma_{ij}-E_{|ij})
   \end{array}
 \right),
\end{equation}
and following combination of gauge-dependent variables are shown
to be unchanged under gauge transformations:
\begin{eqnarray}
\Phi &=& \phi+\frac 1a [a(B-E')]',\\
\Psi &=& \psi-\frac {a'}a (B-E'),\\
\delta\varphi^{(gi)} &=& \delta\varphi+{}^{(0)}\varphi'(B-E'),\\
\delta\xi^{(gi)} &=& \delta\xi+{}^{(0)}\xi'(B-E').
\end{eqnarray}
Introducing a variable
\begin{equation}
\Theta := \Psi+\frac \lambda {2a^2}\{\frac 12({\cal H}^2 + {\cal K})
 \delta\xi^{(gi)}-{\cal H}\xi'\Psi\},
\label{eqn:deftheta}
\end{equation}
scalar part of the perturbed Einstein equation (\ref{eqn:pertEin}) gives 
two independent equations:
\begin{eqnarray}
&&\nabla^2\Theta 
+\left[\frac{3\lambda\xi'}{4a^2\alpha}({\cal H}^2+{\cal K})-3{\cal H}
\right]\Theta'\nonumber\\
&&{}+\left[-6{\cal H}'-12{\cal H}^2-6{\cal K}-\frac{3\lambda\xi'}{2a^2\alpha}
{\cal H}(2{\cal H}'+{\cal H}^2+{\cal K})+\frac{3\lambda^2\xi'{}^2}
{8a^4\alpha^2}({\cal H}^2+{\cal K}){\cal H}'\right]\Theta\nonumber\\
&&{}-\frac{3\lambda^2\xi'}{16a^4\alpha}({\cal H}^2+{\cal K})^2
\delta\xi^{(gi)}{}'-\frac 12 \varphi'\delta\varphi^{(gi)}{}'\nonumber\\
&&{}+({\cal H}^2+{\cal K})\left[\frac{3\lambda}{4a^2}{\cal H}'+\frac{3\lambda}
{4a^2\alpha}(3{\cal H}'+2{\cal H}^2+2{\cal K})-\frac{3\lambda^3\xi'{}^2}
{32a^6\alpha^2}({\cal H}^2+{\cal K}){\cal H}'\right]\delta\xi^{(gi)}=0
\label{eqn:sp1}\\
&&\Theta'+\left[
{\cal H}+\frac{\lambda\xi'}{4a^2\alpha}(4{\cal H}'+3{\cal H}^2+3{\cal K})
\right]\Theta \nonumber\\
&&-\frac{3\lambda^2\xi'}{16a^4\alpha}({\cal H}^2+{\cal K})(2{\cal H}'
+{\cal H}^2+{\cal K})\delta\xi^{(gi)}-\frac 12\varphi'\delta\varphi^{(gi)}=0
\label{eqn:sp2}
\end{eqnarray}
where $\alpha$ is
\begin{equation}
\alpha = 1-\frac 12 \lambda{\cal H}\xi'.
\end{equation}
Eliminating $\delta\varphi^{(gi)}$ from (\ref{eqn:sp1}) and (\ref{eqn:sp2}),
we can obtain one wave function for $\Theta$. However, for finding numerical
solutions it is more convenient to directly integrate (\ref{eqn:sp1}) and
(\ref{eqn:sp2}). Decomposing $\Theta$ into Fourier mode and using
(\ref{eqn:2xi}) for $\xi$, we obtain the numerical solutions shown in fig.2.
Although the amplitude of the perturbation is somewhat enhanced near the peak 
of Hubble parameter, there is no growing mode appearing in the metric 
perturbation.

\input epsf.sty
\begin{figure}
% \vspace*{0cm}
% \hspace*{0cm}
\epsfxsize=12cm
\epsfysize=9cm
\epsffile{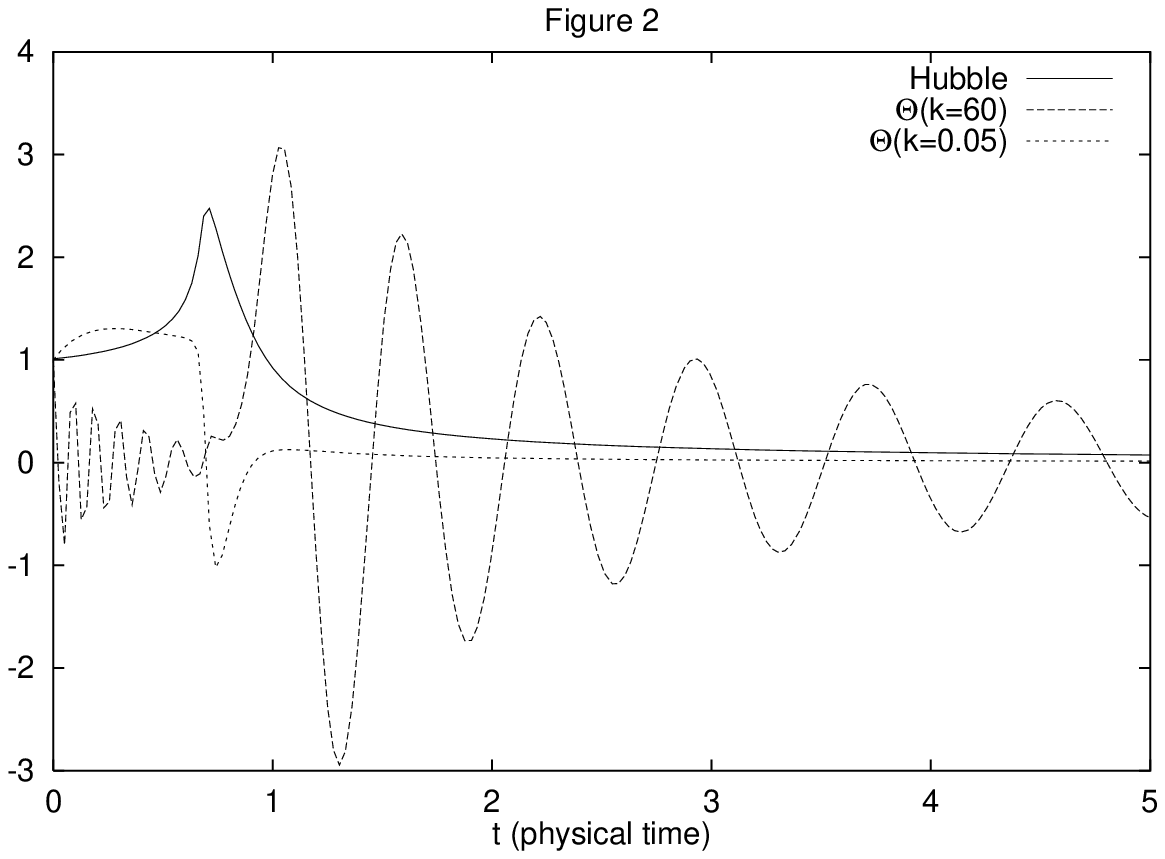}
\caption{Behavior of scalar perturbation. Perturbations with different wave
numbers are shown here, and there is no growing mode.}
\epsfxsize=12cm
\epsfysize=9cm
\epsffile{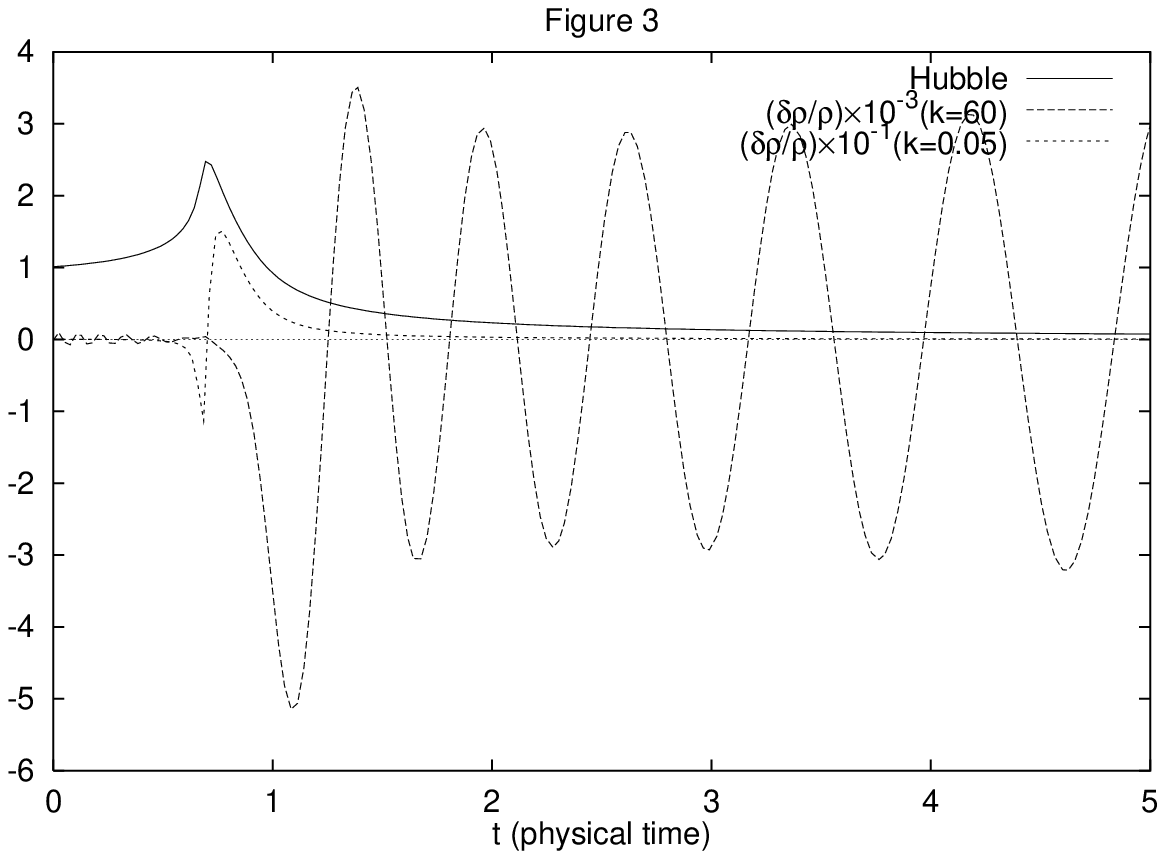}
\caption{Density contrast calculated using the solution shown in Fig.2. 
Small scale density contrasts grow as $\frac{\delta\rho}{\rho}
\propto t^{1/3}$ due to the free scalar field in the 
Friedmann universe.}
\end{figure}

Fig. 3 shows the effective density contrast defined by:
\begin{equation}
\frac{\delta \rho}{\rho} := \frac{\delta^S G^0{}_0}{G^0{}_0}.
\end{equation}
We notice that the small scale effective density contrast grows in Friedmann 
phase. In late time in Friedmann phase, 
the effect of the Gauss-Bonnet term becomes negligible and the evolution of the
density contrast can be considered as in the ordinary Friedmann universe. 
Thus the growth of the density contrast is the consequence of the free scalar 
field remaining in the Friedmann universe. This result is analytically 
understood as follows. Putting $\lambda=0$ in equations (\ref{eqn:sp1}) and
(\ref{eqn:sp2}), we have a equation for $\Theta$:
\begin{equation}
\Theta''-\nabla^2\Theta+6{\cal H}\Theta'+(10{\cal H}'+20{\cal H}^2)\Theta=0.
\end{equation}
Using back ground equations (\ref{eqn:bg1})(\ref{eqn:bg2})(\ref{eqn:bg3}) with
$\lambda=0$, and introducing
\begin{eqnarray}
u &=& \frac{a\Theta}{\varphi'}\\
\theta &=& \frac{{\cal H}}{a\varphi'},
\end{eqnarray}
the equation can be written in a simple form\cite{gip}:
\begin{equation}
u''-\nabla^2u-\frac{\theta''}{\theta}u=0
\label{eqn:frwave}
\end{equation}
Regarding the perturbation as plane wave, the solution for this equation can 
be easily found in both large and small wave number limits. For large enough 
wave number 
($k^2\gg \theta''/\theta)$, the solution of the equation(\ref{eqn:frwave}) is
\[u\propto\exp (\pm ik\eta).\]
Transforming this back into the gauge-invariant perturbative variables, for
density contrast we have
\begin{equation}
\frac{\delta\rho}{\rho} \propto a \propto \eta^{1/2} \propto t^{1/3}.
\end{equation}
Thus the density contrast grows like $t^{1/3}$.

\subsection{Vector part}
The perturbed metric for vector perturbation is defined as:
\begin{equation}
 \delta^V g_{\mu\nu}= a^2(\eta)\left(
   \begin{array}{cc}
     0 & -S_i\\
     -S_j & F_{i|j}+F_{j|i}
   \end{array}
 \right)
\end{equation}
Vector function $S_i$ and $F_i$ satisfy $S_i{}^{|i}=F_i{}^{|i}=0$,
and the combination $P_i = S_i+F_i' $
is unchanged under the gauge transformation.
Vector part of the perturbed Einstein-like equation (\ref{eqn:pertEin}) gives 
2 non-trivial equations:
\begin{equation}
(1-\frac \lambda {2a^2}{\cal H}\xi')(\frac 12 \nabla^2+{\cal K})P_i=0,
\label{eqn:vec1}
\end{equation}
\begin{equation}
\{(1-\frac \lambda{2a^2}{\cal H}\xi')(P_{i|j}+P_{j|i})\}'
 +2{\cal H}(1-\frac \lambda{2a^2}{\cal H}\xi')(P_{i|j}+P_{j|i})=0.
\label{eqn:vec2}
\end{equation}
Equation (\ref{eqn:vec1}) shows the spatial behavior of the vector
perturbation, indicating the fluctuation of curvature scale. 
Equation (\ref{eqn:vec2}) can be easily integrated to give
\begin{equation}
P_{i|j}+P_{j|i}\propto a^{-2}(1-\frac \lambda {2a^2}{\cal H}\xi')^{-1}
=\frac 1{a^2\alpha}.
\end{equation}
which indicates that the vector perturbation decreases as the universe expands.

\subsection{Tensor part}
The tensor part of the metric perturbation is defined as:
\begin{equation}
 \delta^T g_{\mu\nu}=a^2(\eta)\left(
   \begin{array}{cc}
     0 & 0\\
     0 & h_{ij}
   \end{array}
 \right),
\end{equation}
which satisfies the constraint $h^i{}_i=0$, $h_{ij}{}^{|j}=0$.
Tensor part of (\ref{eqn:pertEin}) gives one non-trivial equation for
i-j part:
\[\{(1-\frac \lambda{2a^2}{\cal H}\xi')h^i{}_j{}'\}'
 +2{\cal H}(1-\frac \lambda{2a^2}{\cal H}\xi')h^i{}_j{}'
 -(1+\frac\lambda{2a^2}{\cal H}\xi'-\frac\lambda{2a^2}\xi'')
 (\nabla^2-2{\cal K})h^i{}_j=0\]
Expanding $h_{ij}$ with transverse-traceless basis tensors as
\begin{equation}
h_{ij} = h_+ (k,\eta)\mbox{\bf e}_+{}_{ij}(k)+h_\times(k,\eta)\mbox{\bf e}_\times{}
_{ij}(k).
\end{equation}
for each polarization mode the function $h(k)$ satisfies the
equation of motion using the physical time:
\begin{equation}
\ddot h +(3H+\frac{\dot\alpha}{\alpha})\dot h +\frac{2{\cal K}+k^2}{a^2\alpha}
(1-\frac\lambda 2\ddot\xi)h =0
\label{eqn:tenp}
\end{equation}
Numerical solutions for $h(k)$ with different wave numbers ,in the same 
background as scalar and vector perturbations, are shown in the fig.4. 
There is an instability in the de Sitter-like phase, which results from 
large positive value of $\ddot\xi$ in the equation (\ref{eqn:tenp}).

\input epsf.sty
\begin{figure}
% \vspace*{0cm}
% \hspace*{0cm}
\epsfxsize=12cm
\epsfysize=9cm
\epsffile{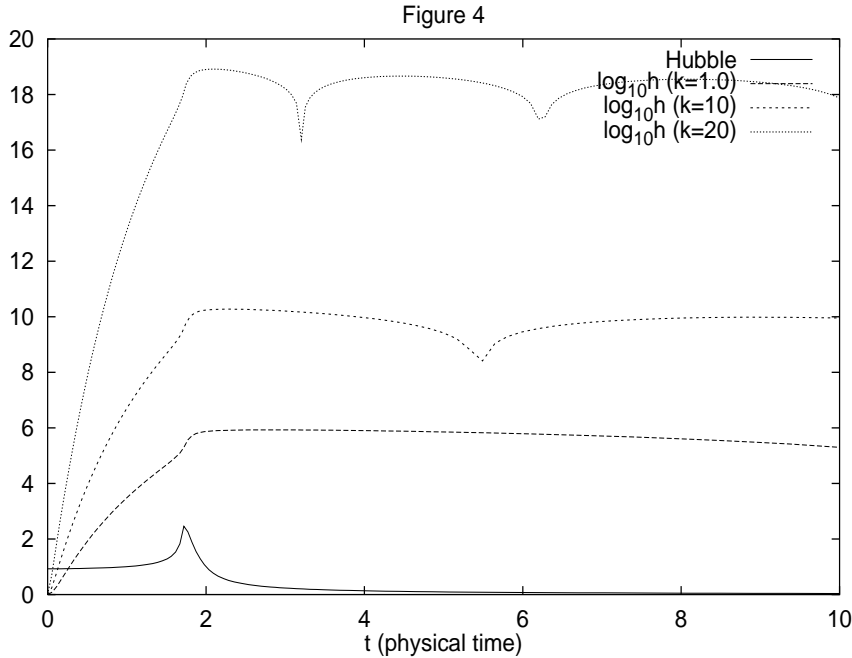}
\caption{Tensor perturbation in the same background. Note that the 
tensor perturbations are plotted with logarithmic scale. Growing mode appears
in the initial de Sitter-like phase. The smaller the perturbation scale is, 
the larger the growth rate becomes.}
\end{figure}

%%%%%%%%%%%%%%%%%%%%%%%%%%%%%%%%%%%%%%%
\section{Conclusion}
%%%%%%%%%%%%%%%%%%%%%%%%%%%%%%%%%%%%%%%
We have made a perturbative analysis of the cosmological model presented by
Rizos and Tamvakis \cite{rt94}, to investigate the stability of the model.
The perturbative equations of motion are solved numerically, and we found that
the system is unstable under tensor perturbation. This instability appears in 
the de Sitter-like stage, and kinetic-driven inflation is thought to be
responsible for the growing mode of the perturbation. Since the form of the
function $\xi$ coming from superstring effective action gives different 
asymptotic behavior of $H$, from this calculation alone we cannot conclude
that this instability continues from the infinite past. But anyway, even 
with the modulus-geometry coupling of the form (\ref{eqn:ssxi}), there exists
a tensor-part instability at least right before the Hubble parameter peak.
Taking this model more seriously, we have to point out some problems.
First of all, we have to check the validity of the effective action 
(\ref{eqn:ss1leAct}). Since the inclusion of the 1-loop effect predicts
somewhat different character of the solution, it is certainly possible that 
the higher loop correction might change the nature of the cosmological model 
drastically. In this analysis we have ignored the effects of dilaton, but
of course we must take these into acount. Viewing this model as a realistic 
model of our universe, we also have to find a mechanism to create the 
ordinary matter, i.e. there have to be something like reheating in the 
inflationary universe. 

%%%%%%%%%%%%%%%%%%%%%%%%%%%%%%%%%%%%%%%
\section*{Acknowledgements}
%%%%%%%%%%%%%%%%%%%%%%%%%%%%%%%%%%%%%%%
We would like to thank D. Wands and H. K. Jassal for useful discussions.

%%%%%%%%%%%%%%%%%%%%%%%%%%%%%%%%%%%%%%%


\begin{thebibliography}{99}
\bibitem{sstr}
 M. Green, J. Schwarz and E. Witten, {\it Superstring Theory, vol. 1\&2},
Cambridge Univ. Press(1987).\\
 There are many review articles on e-print. e.g. 
E. Kiritsis, hep-th/9709062.

\bibitem{pbb}
 M. Gasperini and G. Veneziano, Astropart. Phys. {\bf 1}(1993) 317, 
 Mod. Phys. Lett. {\bf A8}(1993) 3701, Phys. Rev. {\bf D50}(1994) 2519

\bibitem{inflation}
 A. Guth, Phys. Rev. {\bf D23}(1981) 347;
 K. Sato, Mon. Not. R. Astr. Soc{\bf 195}(1981) 467;
 K. Olive, Phys. Rep. {\bf 190}(1990) 307.

\bibitem{sfduality}
 Veneziano, Phys. Lett. {\bf B265}(1991) 287

\bibitem{graceful}
 M. Gasperini, M. Maggiore and G. Veneziano, Nucl. Phys. {\bf B494}(1997) 315;
 R. Brunstein and R. Madden, hep-th/9702043.

\bibitem{turner97}
 M.S.Turner and E. J. Weinberg, Phys. Rev. {\bf D56}(1997) 4604

\bibitem{hwang96}
 J. Hwang, hep-th 9608041

\bibitem{nonsingular}
 A. A. Starobinsky, Phys. Lett. {\bf B91}(1980) 99;
 V. Mukhanov and R. Brandenberger, Phys. Rev. Lett. {\bf 68}(1992) 1969;
 R. Brandenberger, V. Mukhanov and A. Sornborger, Phys. Rev. {\bf D 48}(1993)
  1629.

\bibitem{initsing}
 R. Penrose, Phys. Rev. Lett. {\bf 14}(1965) 57;
 S. Hawking, Proc. R. Soc. London {\bf A 300}(1967) 182;
 S. Hawking and R. Penrose, Proc. R. Soc. London {\bf A314}(1970) 529;
 Also see, e.g. 
 S. W. Hawking and G. F. R. Ellis, {\it The Large Scale Structure of 
Space-time}, Cambridge Univ. Press(1973).

\bibitem{1loop}
 Antoniadis, Gava and Narain, Phys. Lett. {\bf B283}(1992) 209;
 Nucl. Phys. {\bf B393}(1992) 93
 
\bibitem{art94}
 Antoniadis, Rizos and Tamvakis, Nucl. Phys.{\bf B415}(1994) 497

\bibitem{maeda96}
 R. Easther and K. Maeda, Phys. Rev. {\bf D54}(1996) 7252

\bibitem{rt94}
 Rizos and Tamvakis, Phys. Lett. {\bf B326}(1994) 57

\bibitem{gip}
 J. Bardeen, Phys. Rev. {\bf D22}(1980) 1882;
 Mukhanov, Feldman and Brandenberger, Phys. Rep.{\bf 215}(1992) 203;
 H. Kodama and M. Sasaki, Prog. Theor. Phys. Suppl.{\bf 78}(1984) 1.
\end{thebibliography}
\end{document}